\documentclass[11pt,sort&compress]{article}
\usepackage{amsmath} 
\usepackage{mathrsfs}
\usepackage{amssymb} 
\usepackage{graphicx}
\usepackage{bm} 
\usepackage[compress]{cite}
\usepackage[margin=.8in]{geometry}
\usepackage{amsfonts,amsmath,amssymb}
\usepackage{fancyhdr}
\usepackage{dcolumn}
\usepackage{xcolor}
\usepackage{mathrsfs}
\usepackage{amsfonts}
\usepackage{varioref}
\usepackage{textcomp}
\usepackage{url}
\labelformat{figure}{Fig.~#1} 
\labelformat{section}{Section #1} 
\labelformat{subsection}{Section #1} 
\labelformat{subsubsection}{Section #1}
\labelformat{subsubsubsection}{Section #1}
\labelformat{equation}{Eq.~(#1)} 
\labelformat{inequality}{inequality.~(#1)}
\labelformat{subfigure}{Fig.~\thefigure#1} 
\labelformat{table}{Table~#1} 
\labelformat{appendix}{Appendix #1}
 \RequirePackage[colorlinks,citecolor=cyan,urlcolor=blue,linkcolor=blue]{hyperref}
\allowdisplaybreaks
\usepackage{hyperref}
\usepackage{filecontents}

\begin{document}
\title{\textbf{Massive Scalar Perturbation of Extremal Rotating Braneworld Black hole: Superradiant Stability Analysis}}
\date{}
\author{Shauvik Biswas\footnote{\href{mailto:intsb6@iacs.res.in}{intsb6@iacs.res.in}}$~^{}$\footnote{\href{mailto:shauvikbiswas2014@gmail.com}{shauvikbiswas2014@gmail.com}}$~^{}$
\\
$~^{}${\small{School of Physical Sciences}}\\
{\small{Indian Association for the Cultivation of Science, Kolkata-700032, India}}}
\maketitle
\begin{abstract}
We analyse the superradiant stability of braneworld extremal Kerr and Kerr-Newman black holes under massive scalar perturbation. These black hole solutions differ from their four dimensional counterpart  by the presence of a tidal charge, which unlike electric charge, can take both positive and negative values and carries the signature of extra dimensions. We consider the perturbation of the brane geometry by a massive scalar field. From the radial equation of motion of the scalar field, we have found the effective potential felt by the field. For superradiant stability, there should not be any trapping well of the effective potential outside of the horizon. Using this condition we have specified the parameter space of superradiant stability which depends on parameters of both black hole and perturbation. In the case of Kerr-like extremal braneworld black holes, we have obtained  a bound on tidal charge for superradiant stability under a massive scalar perturbation. Similarly, for slowly rotating Kerr-Newman like extremal braneworld black holes, we have obtained a lower bound of the product of mass to charge ratio of the black hole and scalar field, for which the system is superradiantly stable.
\end{abstract}
\section{Introduction}\label{section-1}
Black holes are solutions to Einstein's equation in different scenarios\cite{padmanabhan2010gravitation,d1992introducing,chandrasekhar1998mathematical}. These solutions play important role in high energy physics and astrophysics. The fundamental role of these solutions makes it highly important to study the nature of their stability under linear perturbations of different spins. In this context one considers a test wave impinging on a black hole, a part of which is reflected by the effective potential (felt by it) and scattered to infinity. The transmitted part is swallowed by the black hole. Thus, it is expected that the amplitude of the outgoing wave is  smaller than that of the incident wave.  It was shown that the Schwarzschild black hole is stable under all kinds of massless perturbation \cite{Press:1973zz,Regge:1957td,Vishveshwara:1970cc,Zerilli:1971wd}. On the other hand, the stability issue of the rotating black holes is more complicated due to the presence of ergoregion outside the event horizon. In this region the conserved energy of causal matter becomes negative. One can utilize this fact to extract the rest energy and rotational energy of the black hole\cite{PhysRevLett.25.1596,penrose1971extraction,Arvanitaki:2009fg}. It is shown \cite{Brito:2015oca,Press:1972zz} that when a bosonic wave impinges on a Kerr-like black hole, the reflected wave can be amplified if the so-called superradiance condition \cite{padmanabhan2010gravitation} is satisfied. In this case, the wave near the horizon becomes outgoing \cite{Press:1972zz} and extracts the rotational energy of the black hole from the ergosphere. If this process continues to occur then a Kerr-black hole finally becomes a Schwarzschild black hole. Due to the Pauli exclusion principle, such phenomena are absent in the case of fermionic scattering\cite{Brito:2015oca}.

Press and Teukolsky \cite{Press:1972zz} had considered massless scalar perturbation of Kerr black hole and proposed a feedback mechanism of the superradiantly amplified scattered wave by surrounding the black hole with a reflecting mirror. Then the amplified scattered wave can not escape to infinity and is reflected by the mirror. This wave is then amplified again. In this way, the total extracted energy should grow indefinitely until radiation pressure destroys the system\cite{Press:1972zz,Cardoso:2004nk}. This is called a "black hole bomb". It was also shown \cite{Damour:1976kh} that such a mirror can be realized by using a charged massive scalar field propagating in the Kerr-Newman background. In this case, the effective potential of the massive scalar field has a local minima \cite{Damour:1976kh,Furuhashi:2004jk}, that is outside the horizon we have trapping well and the scalar field can be trapped inside this well instead of letting it escape to infinity\cite{Arvanitaki:2009fg,Hod:2009cp}. Due to the above mentioned feedback mechanism, the scalar field plus black hole system become superradiantly unstable. Various work on instability growth rate had been done in the range $M\mu\ll1$ \cite{Furuhashi:2004jk,Detweiler:1980uk} and $M\mu\gg1$ \cite{Zouros:1979iw}, where M and $\mu$ are the mass of the black hole and scalar field respectively. It is proposed \cite{Hod:2009cp} that this instability growth rate is maximum for extremal black holes, for which the inner and outer horizons coincide. 

In this paper, we are going to study the above mentioned massive scalar perturbation in the context of rotating braneworld black holes. Braneworld black holes are natural candidates to study quantum black holes \cite{Dey:2020lhq,Dey:2020pth,Soleimani:2016mfh,Emparan_2000}. Recently these solutions are extensively used in ECO phenomenology\footnote{Recently an interesting work\cite{Hod:2017cga}analytically shows that highly compact rotating ECO with reflective boundary condition will be marginally stable under static massless scalar perturbation only for discrete spectra of critical radii.} \cite{Dey:2020pth}, tidal heating of black holes \cite{chakraborty2021tidal}, optical properties of black holes \cite{zhang2020optical,PhysRevD.96.084017} and in some other interesting  cases \cite{bronnikov2020echoes,Neves:2020doc,mishra2021constraining,Banerjee:2019nnj} . These black hole solutions differ from their counterpart in general relativity by the presence of a tidal charge, unlike electric charge tidal charge can take negative values, which has no analogue in GR. This charge depends on the electric part of the bulk Weyl tensor and carries the effect of extra dimensions. An immediate consequence of tidal charge is that in the case of extremal Kerr-like black holes we do not have a=M, where a is the black hole spin parameter and M is the mass parameter. Moreover, $a>M$ is allowed, which is also absent in GR. As mentioned before, some earlier works \cite{Zouros:1979iw,Detweiler:1980uk,Hod:2009cp} have shown that the superradiant instability growth rate of massive scalar perturbation of rotating black holes is maximum for extremal black holes. Here we will do a superradiant stability analysis for rotating braneworld black holes under massive scalar perturbation. That is, we will try to determine a parameter space where braneworld Kerr-like (or Kerr-Newman like) black holes are superradiantly stable under massive scalar (or charged massive scalar) perturbations. To do the analysis we will follow the formalism presented in \cite{Hod:2012zza,Lin:2021ssw,Xu:2020fgq,Huang:2019xbu}. As stated before, superradiant instability happens when there exists a local minima (that is trapping well) of effective potential outside the horizon. So for stability under superradiance there should exist only one maxima of effective potential outside the horizon. This is the key argument of our analysis. In this analytic treatment, we will consider only the real frequencies of the scalar wave. But the conclusions can be extended to nearly real frequencies as in \cite{Damour:1976kh,Dolan:2007mj,Zouros:1979iw}. Like \cite{Damour:1976kh} we will assume that both the superradiance condition and the bound state condition hold simultaneously. In the case of Kerr-like extremal braneworld black holes, we will obtain a bound on tidal charge for superradiant stability under a massive scalar perturbation. Similarly, for Kerr-Newman like slowly rotating extremal braneworld black holes, we will obtain a lower bound of the product of mass to charge ratio of the black hole and scalar field for which the system is superradiantly stable. 

This paper is organized as follows: In \ref{section-2} a brief review of the braneworld scenario and the solution of the effective gravitational equation on brane will be given. In \ref{section-3} we will demonstrate the massive scalar perturbation of braneworld extremal Kerr-like black holes. Then taking both the bound state condition and superradiance condition we will find a parameter region where the system is superradiantly stable. In \ref{section-4} we will do the same analysis for charged massive scalar perturbation of Kerr-Newman like extremal braneworld black hole. In \ref{section-5} a discussion of this formalism will be given. \\
\emph{Notations and Conventions:} In our calculations we will set $c=G=1$. The lowercase Roman indices $a,b,c,\ldots$ will denote the four dimensional braneworld  spacetime coordinates and run over 0,1,2,3. On the other hand uppercase Roman indices $A,B,C,\ldots$ will denote the five dimensional bulk spacetime coordinates and run over 0,1,2,3,4. In both the cases 0 denotes the time coordinate. We will follow the positive signature convention for our metric, for which the Minkowski metric in four dimension takes the form $\textrm{diag}(-1,+1,+1,+1)$.

  \section{A brief introduction to the braneworld scenario and solution of the effective gravitational equation on the brane}\label{section-2}
 In the braneworld model, we describe four dimensional  spacetime as embedded in five dimensional spacetime, with a spacelike extra dimension. The non-gravitational physics of matter fields are confined in the four dimensional space time; it is called brane. On the other hand, only gravity can probe the entire five dimensional space-time, known as bulk. People would like to know whether one can have a brane localized black hole. To answer  this one has to solve the effective gravitational equation on the brane, which can be done by the following steps: i) we start by assuming that bulk spacetime satisfies five dimensional Einstein's equation. ii) then we introduce projector $h^{A}_{B}=\delta^{A}_{B}-n^{A}n_{B}$ to the brane, where $n^{A}$ is normal to brane hypersurface, which satisfies $n^{A}n_{A}=1$. iii) Ultimately we use this projector and Gauss-Codazzi relations to derive the Einstein tensor on the brane and thereby determining the effective equation. Considering a vacuum brane, one can get\cite{Shiromizu:1999wj,Dadhich:2000am} \begin{align}\label{vac-brane}
 {}^{(4)}G_{ab}+E_{ab}=0 \quad.
 \end{align}
  In this equation ${}^{(4)}G_{ab}$ is the induced Einstein tensor on the brane and $E_{ab}\equiv C_{PQRS}n^{P}e^{Q}_{a}n^{R}e^{S}_{b}$ is the electric part of the bulk Weyl tensor $C_{PQRS}$. Also $e^{A}_{a}\equiv(\partial x^{A}/\partial y^{a})$, where $x^{A}$ and $y^{a}$ are the brane and bulk the coordinates respectively. $e^{A}_{a}$'s are tangent to brane hypersurface i.e.; $e^{A}_{a}n_{A}=0$.\\
  The static spherically symmetric solution of the above equation was first derived in \cite{Dadhich:2000am}, then using Kerr-Schild ansatz \cite{Bini:2014nga,Kerr:2008zz,Boyer:1966qh} it was generalized to the rotating case \cite{Aliev:2005bi}. Using the Boyer-Lindquist coordinate system, the metric which describes a rotating braneworld black hole with mass $M$ and angular momentum $J\equiv aM$ can be written as,
\begin{align}\label{Kerr-brane}
ds^{2}=-\frac{\Delta}{\rho^{2}}(dt-a{}\sin^{2}\theta d\phi)^2+\frac{\rho^{2}}{\Delta}dr^{2}+\rho^{2}d\theta^{2}+\frac{\sin^{2}\theta}{\rho^{2}}\left[(r^{2}+a^{2})d\phi-adt\right]^{2}~,
\end{align}
where $\Delta\equiv r^{2}+a^{2}-2Mr-Q $  ;  $\rho^{2}\equiv r^{2}+a^{2}\cos^{2}\theta~.$\\


The tidal charge parameter $Q$ depends on the bulk Weyl tensor. It can take both positive and negative values. When $Q$ is negative it is like the case of the Kerr-Newman black hole in GR, on the other hand, if $Q$ is positive, it carries the effect of extra dimensions. The horizons associated with metric in \ref{Kerr-brane} can be obtained by setting $\Delta=0$, this yields,
\begin{align}\label{horizon}
r_{\pm}=M \pm \sqrt{M^2-a^2+Q}~.
\end{align}
Brane localized observers will anticipate $r_{+}=M +\sqrt{M^2-a^2+Q}$ as the horizon of the black hole. Now in case of extremal black holes the two horizon coincides, that is, 
\begin{align}\label{extremal-kerr-hor}
r_{+}=r_{-}=M \quad \text{with} \quad M^{2}=a^{2}-Q~.
\end{align}
If the brane is not vacuum, that is we have a non-vanishing energy momentum tensor $T_{ab}$ of matter fields, then the effective gravitational equation \ref{vac-brane} on brane modifies as\cite{Shiromizu:1999wj,Aliev:2005bi},
\begin{align}\label{not-vac-brane}
{}^{(4)}G_{ab}+E_{ab}=8\pi T_{ab}+\pi_{ab}\quad.
\end{align}
In this equation $\pi_{ab}$ is called "squared" energy momentum tensor; which contains quadratic combinations of $T_{ab}$(for explicit expressions see\cite{Aliev:2005bi,Shiromizu:1999wj}). If a Maxwell field is confined on the brane, then $T_{ab}$ is given by\cite{landau2013classical,padmanabhan2010gravitation,d1992introducing}, \begin{align}
T_{ab}=\frac{1}{4\pi}\left(F_{ac}F^{c}_{b}-\frac{1}{4}h_{ab}F_{cd}F^{cd}\right)~,
\end{align}
where $F_{ab}$ is the electromagnetic field tensor. Using this energy momentum tensor in \ref{not-vac-brane} and then taking Kerr-Schild ansatz\cite{Bini:2014nga,Kerr:2008zz,Boyer:1966qh} one can derive a metric describing rotating black hole on brane having both tidal charge and electric charge \footnote{Note that our notation of electric charge and tidal charge is different from that of the reference given.} \cite{Aliev:2005bi}, which in Boyer-Lindquist coordinate system reads,

\begin{align}
ds^{2}=-(1-H)dt^{2}+\frac{\rho^{2}}{\Delta_{l}}(1+\delta)dr^{2}+\rho^{2}d\theta^{2}+(r^{2}+a^{2}+
Ha^{2}\sin^{2}\theta)\sin^{2}\theta d\phi^{2}~\\
\nonumber
-2\delta(dt-a\sin^{2}\theta d\phi)dr-2aH\sin^{2}\theta dt d\phi~,
\end{align}
where
\begin{align}
H=\frac{2Mr+Q-\beta^{2}}{\rho^{2}}-\frac{l^{2}\beta^{4}}{\rho^{2}}h~,
\end{align}
\begin{align}
\Delta_{l}=r^{2}+a^{2}+\beta^{2}-2Mr-Q+l^{2}\beta^{4}h_{0} \quad ; \quad \delta=-\frac{l^{2}\beta^{4}}{\Delta_{l}}(h-h_{0})~,
\end{align}
\begin{align}
h=\frac{1}{8a^{4}\cos^{4}\theta}\left[2+\frac{r^{2}}{\rho^{2}}+\frac{3r}{a\cos\theta} tan^{-1}\left(\frac{r} { a\cos\theta}\right)\right]~.\quad 
\end{align}
Here $l$ is the AdS curvature radius and $h_{0}$ is the value of $h$ at fixed angle $\theta_{0}$. One can expand $H$ and $h$ in powers of $a$ and keep corrections up to second order to yield,
\begin{align}
\quad h-h_{0}=-\frac{a^{2}}{14r^{6}}(\cos^{2}\theta-\cos^{2}\theta_{0})~,
\end{align}
\begin{align}
H=\frac{2M}{r}-\frac{\beta^{2}-Q}{r^2}-\frac{l^{2}\beta^{4}}{20r^{6}}-\frac{a^{2}\cos^{2}\theta}{r^{2}}\left(\frac{2M}{r}-\frac{\beta^{2}-Q}{r^2}-\frac{17}{140}\frac{l^{2}\beta^{4}}{r^{6}}\right)~.
\end{align}
 Now since the terms of $\mathcal{O}\left(\frac{1}{r^{6}}\right)$ are highly suppressed than the other terms, so we can approximate the metric in  the form of \ref{Kerr-brane} with $\Delta\equiv r^{2}+a^{2}+\beta^{2}-2Mr-Q $, where $\beta$ is the electric charge. 
In this case, the extremal black hole is characterized by the condition: \begin{align}
r_{+}=r_{-}=M \quad \text{with} \quad M^{2}=a^{2}+\beta^{2}-Q~.
\end{align}
In the next section, we are going to study the massive scalar perturbation of the brane geometry, where the extra dimension will be kept unperturbed.
\section{Massive Scalar perturbation of Braneworld Extremal Kerr Black Hole }\label{section-3}
The dynamics of a massive scalar perturbation $\Psi$ in the  Kerr background \ref{Kerr-brane} is described by the covariant Klein-Gordon equation\cite{Zouros:1979iw,Detweiler:1980uk,Rowan:1977zg},
\begin{align}\label{wf-1}
(\nabla^{a}\nabla_{a}-\mu^{2})\Psi=0~,
\end{align}
where $\mu$ is the mass of the scalar field and $\nabla_{a}$ represents the covariant derivative with respect to the background brane metric.

 Since the background spacetime exhibits two killing vectors field,  $(\partial/\partial t)^{a}$ and $(\partial/\partial \phi)^{a}$, so we can decompose the solution of the above differential equation as, \begin{align}\label{wf-decomposition}
\Psi(t,r,\theta,\phi)=\sum_{lm}R_{lm}(r)S_{lm}(\theta)e^{im\phi}e^{-i\omega t}~,
\end{align}
 here $l$ and $m$ respectively represent the spheroidal harmonics index and azimuthal  harmonics index of the mode. $R_{lm}(r)$ is the radial wave function and $S_{lm}(\theta)$ is the angular wave function. 
$S_{lm}(\theta)$ satisfies the spheroidal harmonic equation\cite{PhysRevLett.29.1114,Hartman:2009nz,PhysRevLett.100.121101,Fiziev:2009kh,Hod:2015cqa}:
\begin{align}\label{anguler-per-kerr}
\frac{1}{\sin\theta}\frac{d}{d\theta}\left(\sin\theta\frac{d S_{lm}}{d\theta}\right)+\left[K_{lm}+(\mu^{2}-\omega^{2})a^{2}\sin^{2}\theta-\frac{m^{2}}{\sin^{2}\theta}\right]S_{lm}=0~,
\end{align}
here $K_{lm}$ is the eigenvalue of the differential operator with $S_{lm}(\theta)$ as eigenfunction. Lower bound of $K_{lm}$ is given by \cite{Hod:2015cqa,Bardeen:1972fi,Hod:2016iri}: 
\begin{align}\label{lower-bound-eigenvalue-K}
K_{lm}> m^{2}-a^{2}(\mu^{2}-\omega^{2}).
\end{align}
On the other hand $R_{lm}(r)$ satisfies the differential equation\cite{Carter:1968ks,PhysRevLett.29.1114,Hartman:2009nz,PhysRevLett.100.121101,Fiziev:2009kh}, 
\begin{align}\label{K-radial}
\Delta\frac{d}{dr}\left(\Delta\frac{dR_{lm}}{dr}\right)+U_{K}R_{lm}=0~,
\end{align}
where \begin{align}\label{Takulsky-like-potential}
U_{K}=\{\omega(r^{2}+a^{2})-am\}^{2}+\Delta\{2am\omega-\mu^{2}(r^{2}+a^{2})-K_{lm}\}.
\end{align}
We define tortoise coordinate as
\begin{align}\label{tortoise}
\frac{dr_{*}}{dr}=\frac{r^{2}+a^{2}}{\Delta}~.
\end{align}
By defining a new radial function \cite{Press:1973zz} $\tilde{R}_{lm}\equiv\sqrt{r^{2}+a^{2}}R_{lm}$ we can bring \ref{K-radial} in a form 
\begin{align}
\frac{d^{2}\tilde{R}_{lm}}{dr_{*}^{2}}+\tilde{U}_{K}\tilde{R}_{lm}=0~,
\end{align}
where
\begin{align}\label{tilde U}
\tilde{U}_{K}=\frac{1}{(r^{2}+a^{2})^{2}}\left[(r^{2}+a^{2})^{2} \left(\omega-\frac{ma}{(r^{2}+a^{2})}\right)^{2}+\Delta(r^{2}+a^{2})\left(-\mu^{2}-\frac{K_{lm}}{(r^{2}+a^{2})}+\frac{2am\omega}{(r^{2}+a^{2})}\right)\right]-\frac{\Delta P(r)}{(r^{2}+a^{2})^{2}}
\end{align}
with $P(r)\equiv\frac{[\{r\Delta^{\prime}+\Delta\}(r^{2}+a^{2})^{\frac{3}{2}}-3r^{2}\Delta\sqrt{r^{2}+a^{2}}]}{(r^{2}+a^{2})^\frac{5}{2}}$.

Considering asymptotic and near horizon behaviour of $\tilde{U}_{K}$ and demanding exponentially decaying bound state solution at spatial infinity we impose the following boundary condition on $\tilde{R}_{lm}$ 
\begin{align}
\tilde{R}_{\ell m}(r)\sim
    \begin{cases} \label{asymp_ampli}
     e^{-\sqrt{\mu^{2}-\omega^{2}}r_{*}} & \text{for}\ \quad r_{*}\to \infty
       \\\\
     e^{-i(\omega-m\Omega_{H})r_{*}} & \text{for} \quad r_{*}\to -\infty~,
    \end{cases} 
\end{align}\label{Boundary-kerr} 
where $\Omega_{H}=\frac{a}{r^{2}_{+}+a^{2}}=\frac{a}{2a^{2}-Q}$ is the angular velocity of the horizon. 
Here we have used the fact, that near the horizon we should have only ingoing wave (as seen by the co-moving observers). 

From the above, it is clear that to get exponentially decaying bound state at spatial infinity we need 
\begin{align}\label{exponentially-bound-1}
\mu^{2}>\omega^{2}.
\end{align}
In case of Kerr-like braneworld black holes the superadiance condition\cite{Brito:2015oca,PhysRevD.7.949,Dey:2020pth}is given by   
\begin{align}\label{superradiance condition-kerr}
0<\omega< m\Omega_{H}\quad (m>0).
\end{align}

\subsection{Superradiant Stability Analysis}\label{Stabilit-K}
In order to find superradiantly stable parameter space of the braneworld extermal Kerr black hole, we define a new radial function \cite{Hod:2012zza} $\psi_{lm}=\sqrt{\Delta}R_{lm}$. Using this radial function in \ref{K-radial} we get 
\begin{align}
\frac{d^{2}\psi_{lm}}{dr^{2}}+(\omega^{2}-V_{K})\psi_{lm}=0~,
\end{align}
where \begin{align}
V_{K}&=\omega^{2}-\frac{U_{K}}{\Delta^{2}}\\
&=\omega^{2}-\frac{\{\omega(r^{2}+a^{2})-am\}^{2}}{\Delta^{2}}-\frac{\{2am\omega-\mu^{2}(r^{2}+a^{2})-K_{lm}\}}{\Delta}~.
\end{align}
Here we have used the fact, that for extremal black holes  $M^{2}=a^{2}-Q$.\\

The asymptotic and near horizon behaviour of the above potential is given by 
\begin{align}
V_{K}\sim
    \begin{cases} \label{asymp_V_k}
     \mu^{2}+\frac{2M}{r}(\mu^{2}-2\omega^{2})+\mathcal{O}(\frac{1}{r^2}) & \text{for}\ \quad r\to \infty
       \\\\
     -\infty & \text{for} \quad r\to r_{+}.
    \end{cases} 
\end{align}\label{Asymptotic -near horizon behavoiur}

On the other hand   $V^{\prime}_{K}(r\rightarrow \infty)\rightarrow\frac{2M}{r^2}(-\mu^{2}+2\omega^{2})+\mathcal{O}(\frac{1}{r^3})$. As explained in \ref{section-1}, for superradiant stability under massive scalar perturbation, there should not be any trapping well outside the horizon  for the effective  potential  $V_{K}$\cite{Hod:2012zza,Lin:2021ssw,Xu:2020fgq}. That is, it can have only one maxima outside the horizon and no local minima. To do such analysis, we compute the derivative of effective potential $V_{K}$ using a new radial coordinate $z=r-M$. In terms of this new radial coordinate, one can write the derivative of $V_{K}$ as,
\begin{align}\label{derivative of potential }
V^{\prime}_{K}(z)&=\frac{z(Az^{3}+Bz^{2}+Cz+D)}{-z^{6}}\equiv\frac{zf(z)}{-z^{6}}~,
\end{align}
where \begin{align}\label{coefficients-Kerr-A}
A&=2M(\mu^{2}-2\omega^{2})~,\\
B&=2K_{lm}+2\mu^{2}a^{2}+2\mu^{2}M^{2}-12\omega^{2}M^{2}-4\omega^{2}a^{2}~,\\\label{coefficients-Kerr-B}
C&=12M\omega m a -12M\omega^{2}(M^{2}+a^{2})~,\\
D&=-4[\omega(M^{2}+a^{2})-ma]^{2}~.\label{coefficients-Kerr-D}
\end{align}
Here we have used the fact, that for extremal black holes $\Delta=z^{2}$. In the above expressions $M^{2}=a^{2}-Q$. \\

 At the position of maxima or minima $V^{\prime}_{K}(r)=V^{\prime}_{K}(z)=0$. Then we can avoid to have any trapping well (local minima) outside the horizon if the equation $V^{\prime}_{K}(z)=0$ has only one positive root and $V^{\prime\prime}_{K}(z)<0$ at that point. That is, one can remove the possibility of getting the local minima outside the horizon by requiring that $V_{K}(r)$ can have only one maxima outside the horizon. To specify the parameter region consistent with this requirement, we analyze the sign of the roots of the $f(z)=0$ equation. This can be done using Vieta's theorem\cite{barnard1959higher}. If $z_{1}$, $z_{2}$ and $z_{3}$ are the three roots of the cubic equation $f(z)=0$, then according to this theorem: 
\begin{align}\label{Vieta-Kerr}
z_{1}+z_{2}+z_{3}=-\frac{B}{A}~,\\
z_{1}z_{2}+z_{2}z_{3}+z_{3}z_{1}=\frac{C}{A}~,\label{product-sum-KR}\\
z_{1}z_{2}z_{3}=-\frac{D}{A}~.\label{Product-Kerr}
\end{align}
From the expression of the coefficients of $f(z)$ it is clear that $D<0$. Now suppose $A<0$, then from \ref{Product-Kerr} we can infer that the product of the roots is negative. In this case either we will have two positive root and one negative root or the three roots will be negative. Now using superradiance condition \ref{superradiance condition-kerr} we can see that $C>0$. Then from \ref{product-sum-KR} it is clear that we can never have three negative roots if $A<0$.
In this case we are left with the possibility of having two positive roots and one negative root. That is, with the condition $A<0$ one can not avoid the possibility of forming trapping well outside the horizon. On the other hand, if we let $A>0$, then the product of the roots become positive. In this case, we can choose our parameters to have only one positive (say $z_{1}$) and two negative roots (say $z_{2}$ and $z_{3}$). So the necessary condition to avoid trapping well is $A>0$(see also \cite{HOD20131489}). In terms of perturbation parameters this gives,
\begin{align}\label{condition-at-infiniy}
\omega^{2}<\frac{\mu^{2}}{2}~.
\end{align}
According to the asymptotic behaviour of the effective potential, this implies $V^{\prime}_{K}(r\rightarrow \infty)<0$. 
So we have reduced our job to a problem of finding the parameter region where $z_{2}<0$ and $z_{3}<0$. A sufficient condition for the negativity of both $z_{2}$ and $z_{3}$ is (see \ref{Vieta-Kerr}) \begin{align}\label{condition-B}
B>0~.
\end{align}
Now using superradiance condition \ref{superradiance condition-kerr} and lower bound of $K_{lm}$ \ref{lower-bound-eigenvalue-K}in the inequality \ref{condition-B} we get \begin{align}\label{condition-from-B}
\mu &>\omega\sqrt{3+\frac{Q}{a^{2}}}~.
\end{align}
It only remains to confirm that $z_{1}(>0)$ is the maxima of $V_{K}(z)$ in this parameter region; that is $V^{\prime\prime}_{K}(z_{1})<0$. The second derivative of the potential $V_{K}(z)$ can be written in a form
\begin{align}
V^{\prime\prime}_{K}(z)=\frac{5f(z)-(3Az^{3}+2Bz^{2}+Cz)}{z^{6}}~.
\end{align}
Since $V^{\prime}_{K}(z_{1})=0$, so that $f(z_{1})=0$. It is easy to see that when suprradience condition is satisfied i.e., $ m>\frac{\omega(M^{2}+a^{2})}{a}$, then $C>0$.  Also by our demand $B>0$. So  $V^{\prime\prime}_{K}(z_{1})<0$ in this parameter region.
\subsection{Bound On Tidal Charge}
To hold the inequality \ref{condition-at-infiniy} (otherwise we can not ensure $A>0$) and \ref{condition-from-B} simultaneously, we must have \begin{align}
Q\geq -a^{2}.
\end{align} 
Also we must have $M^{2}=a^{2}-Q >0$. Combining these two inequalities we get 
\begin{align}\label{Bound On Tidal Charge}
 -a^{2}\leq Q<a^{2}.
\end{align}
This serves as a bound on tidal charge. When the inequalities \ref{superradiance condition-kerr}, \ref{condition-from-B} and \ref{Bound On Tidal Charge} are satisfied, then we can ensure that there is no trapping potential well outside the horizon and the system of scalar perturbation and braneworld extremal Kerr black hole is superradiantly stable. Note that when these three inequalities hold simultaneously, the bound state condition \ref{exponentially-bound-1} and the inequality \ref{condition-at-infiniy} hold automatically. Therefore these three inequalities are sufficient to specify superradiantly stable parameter space.
\subsection{Behaviour of Effective Potential Outside the Horizon }
In this section, we will demonstrate the behaviour of effective potential $V_{K}(r)$ as a function of r for two sets of parameters. This will give a justification of our results.
\begin{enumerate}
\item \textit{Case-1 }. In this case, we choose our parameters such that inequalities \ref{superradiance condition-kerr}, \ref{condition-from-B} and \ref{Bound On Tidal Charge} are satisfied simultaneously. For example we take $a=2 $, $m=1$, $\omega=0.250$, $\mu=0.500$, $M=1$, $Q=3$. As required by our result, there is no potential well outside the horizon \ref{figure:1} and the system is superradiantly stable.
\begin{figure}[h]
\begin{center}
 \includegraphics[scale=0.65]{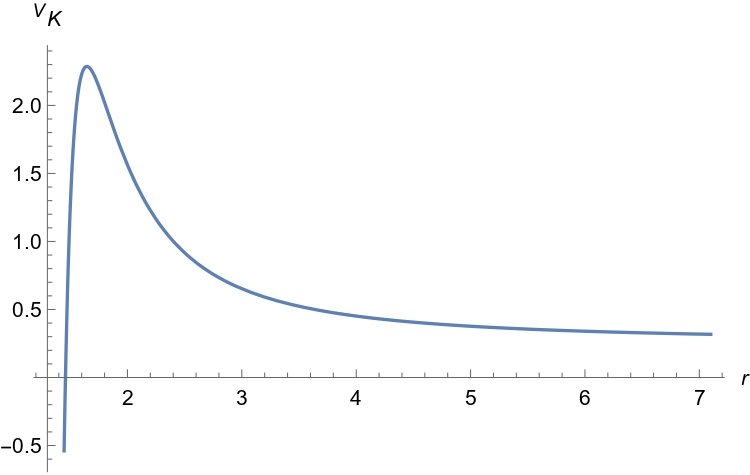}
 \caption{The effective potential $V_{K}(r)$ has no trapping well outside the horizon. The system is superradiantly stable. Here the parameters of perturbation are $\omega=0.250$, $\mu=0.500$ and $m=1$. On the other hand, the parameters of the black hole are chosen as $M=1$, $a=2$ and $Q=3$.}\label{figure:1}
 \end{center}
\end{figure}
\begin{figure}[h]
\begin{center}
\includegraphics[scale=.65]{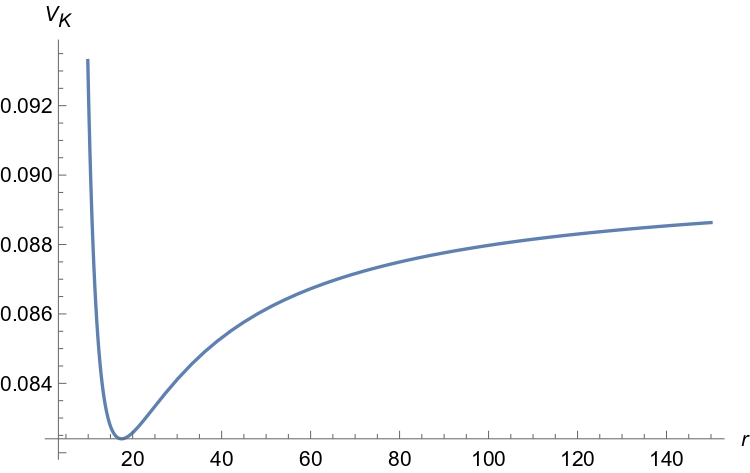}
\caption[]{The effective potential $V_{K}(r)$ has a trapping well outside the horizon. The system is superradiantly unstable. In this case parameters of the perturbation are chosen as $m=2$, $\omega=0.250$ and $\mu=0.300$. On the hand, the black hole parameters are $M=3$, $Q=-5$ and $a=2$.}
\label{figure:2}
\end{center}
\end{figure}
\item  \textit{Case-2}. In this case, we choose our parameters such that they do not satisfy all the above mentioned inequalities.  For example we take $a=2$, $m=2$, $\omega=0.250$, $\mu=0.300$, $M=3$, $Q=-5$. In this case both bound state condition \ref{exponentially-bound-1} and superradiance condition \ref{superradiance condition-kerr} hold but inequality \ref{Bound On Tidal Charge} does not hold. As evident from the \ref{figure:2}, we have a potential well outside the horizon and the system is superradiantly unstable. 
\end{enumerate}

\section{Charged Massive Scalar perturbation of Braneworld Extremal Kerr-Newman Black Hole}\label{section-4} 
As stated in \ref{section-2} in this case the brane geometry is approximately described by the metric \ref{Kerr-brane} with $\Delta\equiv r^{2}+a^{2}+\beta^{2}-2Mr-Q $. The dynamics of a charged massive scalar perturbation $\Psi$ of mass $\mu$ and charge $q$ in the braneworld Kerr-Newman background is given by \cite{PhysRevLett.29.1114,Hartman:2009nz,PhysRevD.94.044036},  
\begin{align}\label{charged-wf}
(\nabla^{a}-iqA^{a})(\nabla_{a}-iqA_{a})\Psi=\mu^{2}\Psi~,
\end{align}
here $\nabla_{a}$ represents covariant derivative in the background geometry and the background electromagnetic vector potential is given by \begin{align}\label{background potential}
A_{a}=\left(\frac{-r\beta}{\rho^2},0,0,\frac{ar\beta sin^{2}\theta}{\rho^2}\right).
\end{align}
Using the same argument presented in \ref{section-3} the solution of the above differential equation can be decomposed as \begin{align}\label{wf-decomposition}
\Psi(t,r,\theta,\phi)=\sum_{lm}R_{lm}(r)S_{lm}(\theta)e^{im\phi}e^{-i\omega t}~,
\end{align}
here $l$ and $m$ receptively represent the spheroidal harmonics index and azimuthal harmonics index of the mode. $R_{lm}(r)$ is the radial wave function and $S_{lm}(\theta)$ is the angular wave function. $S_{lm}(\theta)$ satisfies same the spheroidal harmonic \ref{anguler-per-kerr}. So $K_{lm}$ has the same lower bound \ref{lower-bound-eigenvalue-K}.

$R_{lm}(r)$ satisfies a differential equation of the form \ref{K-radial} but with different potential \cite{PhysRevD.94.044036},    
\begin{align}\label{Kerr-N-radial}
\Delta\frac{d}{dr}\left(\Delta\frac{dR_{lm}}{dr}\right)+U_{KN}R_{lm}=0~,
\end{align}
where 
\begin{align}\label{Takulsky-like-KN-potential}
U_{KN}=\{\omega(r^{2}+a^{2})-am-q\beta r\}^{2}+\Delta\{2am\omega-\mu^{2}(r^{2}+a^{2})-K_{lm}\}~.
\end{align}
One can do a similar analysis as we have done in \ref{section-3} to find the asymptotic and near horizon behaviour of $R_{lm}(r)$. Demanding exponentially decaying bound state solution at spatial infinity and purely ingoing wave solution near the horizon, we can get the following behaviour of $R_{lm}$,   
\begin{align}
R_{\ell m}(r)\sim
    \begin{cases} \label{asymp_kn}
     \frac{e^{-\sqrt{\mu^{2}-\omega^{2}}r_{*}}}{r} & \text{for}\ \quad r_{*}\to \infty
       \\\\
     e^{-i(\omega-\omega_{c})r_{*}} & \text{for} \quad r_{*}\to -\infty~,
    \end{cases} 
\end{align}\label{Boundary-k-n}
where $\omega_{c}=m\Omega_{H}+q\Phi_{H}$. Here $\Omega_{H}=\frac{a}{M^{2}+a^{2}}$ is the angular velocity of the horizon and $\Phi_{H}=\frac{M\beta}{M^{2}+a^{2}}$ is the electric potential of the horizon.
From the above, it is clear that  to get an exponentially decaying bound state at spatial infinity we need 
\begin{align}\label{exponentially-bound}
\mu^{2}>\omega^{2}.
\end{align}
In the case of charged rotating black holes the superradiance condition\cite{Brito:2015oca,PhysRevD.7.949} is, 
\begin{align}\label{superradiance condition-kerr-newman}
0<\omega< \omega_{c}~.
\end{align}
\subsection{Superradiant Stability Analysis}\label{stabiity-KN}
 To analyze the superradiant stability of the braneworld extremal Kerr-Newman black hole we follow the same steps presented in \ref{Stabilit-K}. Using a new radial function \cite{Hod:2012zza} $\psi_{lm}=\sqrt{\Delta}R_{lm}$, one can put the radial equation \ref{Kerr-N-radial} in the form, 
 \begin{align}\label{scr-KN}
\frac{d^{2}\psi_{lm}}{dr^{2}}+(\omega^{2}-V_{KN})\psi_{lm}=0~,
\end{align} 
 where \begin{align}
V_{KN}&=\omega^{2}-\frac{U_{KN}}{\Delta^{2}}\\
&=\omega^{2}-\frac{\{\omega(r^{2}+a^{2})-am-rq\beta\}^{2}}{\Delta^{2}}-\frac{\{2am\omega-\mu^{2}(r^{2}+a^{2})-K_{lm}\}}{\Delta}~.
\end{align}

While deriving \ref{scr-KN} we have used the fact, that for charged extremal black holes $M^{2}=a^{2}+\beta^{2}-Q$.\\
The asymptotic and near horizon behaviour of the effective potential $V_{KN}(r)$ is,
\begin{align}
V_{KN }\sim
    \begin{cases} \label{asymp_V-KN}
     \mu^{2}+\frac{2M(\mu^{2}-2\omega^{2})+2q\beta\omega}{r}+\mathcal{O}(\frac{1}{r^2}) & \text{for}\ \quad r\to \infty
       \\\\
     -\infty & \text{for} \quad r\to r_{+}.
    \end{cases} 
\end{align}\label{Asymptotic -near horizon behavoiur}

On the other hand $V^{\prime}_{KN}(r\rightarrow \infty)\rightarrow\frac{-2M(\mu^{2}-\omega^{2})-2\omega(q\beta-M\omega)}{r^2}+\mathcal{O}(\frac{1}{r^3})$. For superradiant stability under charged massive scalar perturbation, there should not be any trapping well outside the horizon  for the effective  potential  $V_{KN}$ \cite{Hod:2012zza,Lin:2021ssw,Xu:2020fgq}. As in \ref{Stabilit-K}, we can implement this condition by demanding  \begin{enumerate}
\item $V^{\prime}_{KN}(r\rightarrow \infty)<0$~,
\item $V_{KN}(r)$ should have only one maxima outside the horizon.
\end{enumerate} 
If the bound state condition \ref{exponentially-bound} holds then the first condition always holds when 
\begin{align}\label{condition-at-infiniy-KN}
\omega<\frac{q\beta}{M}\quad (q\beta\neq 0)~.
\end{align}
To fulfill the second condition, we compute the derivative of the effective potential $V_{KN}(r)$ using new radial coordinate $z=r-M$, which yields,
  \begin{align}\label{derivative of potential-kn }
V^{\prime}_{KN}(z)&=\frac{z(Ez^{3}+Fz^{2}+Gz+H)}{-z^{6}}\equiv\frac{zh(z)}{-z^{6}}~,
\end{align}
here
\begin{align}
E&=2[\omega(q\beta-M\omega)+M(\mu^{2}-\omega^{2})]~,\label{coe-KN-E}\\
F&=2[\mu^{2}(M^{2}+a^{2})-2\omega\{\omega(M^{2}+a^{2})-qM\beta \}+K_{lm}-(q\beta-2M\omega)^{2}]~,\\
G&=-6(2M\omega-q\beta)\{\omega(M^{2}+a^{2})-am-qM\beta\}~,\\
H&=-4(M^{2}+a^{2})^{2}(\omega-\omega_{c})^{2}~.\label{coe-KN-H}
\end{align}
In the above expressions $M^{2}=a^{2}+\beta^{2}-Q$. As argued in \ref{Stabilit-K}, the effective potential $V_{KN}(r)$ will have only one maxima outside horizon if the equation $h(z)=0$ has only one positive root. If $z_{1}$, $z_{2}$ and $z_{3}$ are the three roots of the cubic equation $h(z)=0$, then according to Vieta's theorem\cite{barnard1959higher}
\begin{align}
z_{1}+z_{2}+z_{3}=-\frac{F}{E}~,\label{sum-kn}\\
z_{1}z_{2}+z_{2}z_{3}+z_{3}z_{1}=\frac{G}{E}~,\\
z_{1}z_{2}z_{3}=-\frac{H}{E}~.\label{product-kn}
\end{align}
From the asymptotic behaviour of the effective potential $V_{KN}(r)$ it is clear that there exists at least one positive root (say $z_{1}$)outside the horizon, that is \begin{align}
z_{1}>0~.
\end{align}
From \ref{coe-KN-E} and using \ref{condition-at-infiniy-KN} and \ref{exponentially-bound} we can see that $E>0$. It is also clear from \ref{coe-KN-H} that $H<0$. So from \ref{product-kn} we see that the product of the roots is positive. Then a sufficient condition for the negativity of both $z_{2}$ and $z_{3}$ (see \ref{sum-kn}) is,\begin{align}\label{condition-F}
F>0~.
\end{align}
Now using the lower bound of $K_{lm}$ \ref{lower-bound-eigenvalue-K} in the inequality \ref{condition-F}
we get, \begin{align}
 \mu^{2}>\frac{\omega^{2}(6M^{2}+a^{2})-6\omega\beta qM + (q^{2}\beta^{2}-m^{2})}{M^{2}}\equiv\frac{y(\omega)}{M^2}~.\label{condition-from-F}
\end{align}
The function $y(\omega)$ represents a parabola of the form $(\omega-\omega_{A})^{2}=4a(y-b)$, 
where 
\begin{align*}
&\omega_{A}=\frac{3q\beta M}{6M^{2}+a^{2}}~,\\
&4a=\frac{1}{6M^{2}+a^{2}}~,\\
&b=\frac{q^{2}\beta^{2}(a^{2}-3M^{2})}{6M^{2}+a^{2}}-m^{2}~,
\end{align*}
here $\omega_{A}$ represents the axis of symmetry of the parabola. Using the inequality $ \frac{1}{6M^{2}+a^{2}}<\frac{1}{6M^{2}}$, one can show that,  
\begin{align}
2\omega_{A}<\frac{q\beta}{M}~.
\end{align}
From this inequality, it is clear that if $0<\omega<\frac{q\beta}{M}$, the maximum value of $y(\omega)$ is given by $y(\frac{q\beta}{M})$. So the sufficient condition to hold the inequality \ref{condition-from-F} is
\begin{align}\label{charge-to-mass}
\mu^{2}&>\frac{y(\frac{q\beta}{M})+m^{2}}{M^2}=\left(\frac{q\beta}{M}\right)^{2}(k^{2}+1)\quad \text{where}\quad k=\frac{a}{M}~.
\end{align}
In terms of tidal charge, the above inequality becomes \begin{align}
\frac{\mu}{q}>\frac{\beta}{\sqrt{a^{2}+\beta^{2}-Q}}\sqrt{\frac{a^{2}}{a^{2}+\beta^{2}-Q}+1}~.
\end{align}
When the above inequality holds together with inequalities \ref{superradiance condition-kerr-newman} and \ref{condition-at-infiniy-KN}, there is no trapping well outside the horizon and our system of charged  massive scalar field and  braneworld extremal Kerr-Newman black is superradiantly stable. When these inequalities are not satisfied, we can not assure the superradiant stability of the system.
\subsection{Behaviour of Effective Potential Outside the Horizon }
In this section, we will demonstrate the behaviour of effective potential $V_{KN}(r)$ as a function of r for two sets of parameters. This will give a justification of our results.
\begin{enumerate}
 \item \textit{Case-1}. In this case we choose our parameters such that inequalities \ref{charge-to-mass}, \ref{condition-at-infiniy-KN} and \ref{superradiance condition-kerr-newman}
are satisfied simultaneously. For example we take $a=0.5$, $m=1$, $\omega=0.280$, $\mu=0.350$, $M=2.961$, $Q=-2.27$, $q=0.4$, $\beta=2.5$. In this case $\frac{M\mu}{q\beta}=1.036>\sqrt{k^{2}+1}=1.014$ and $\omega=0.280 <\frac{q\beta}{M}=0.337$. As required by our results, there is no potential well outside the horizon (see \ref{figure:3}) and the system is superradiantly stable.\\
\begin{figure}[h]
\begin{center}
 \includegraphics[scale=0.65]{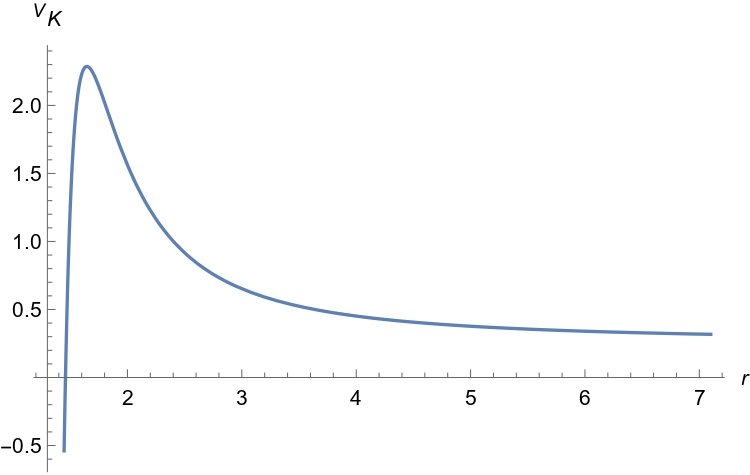}
 \caption[]{The effective potential $V_{KN}(r)$ has no trapping well outside the horizon. Therefore system is superradiantly stable. The parameters of the perturbation are taken as  $m=1$, $\omega=0.280$, $\mu=0.350$, $q=0.4$. The parameters of the black hole are $a=0.5$, $\beta=2.5$, $M=2.961$, $Q=-2.27$.}
 \label{figure:3}
 \end{center}
\end{figure}
\begin{figure}[h]
\begin{center}
 \includegraphics[scale=0.65]{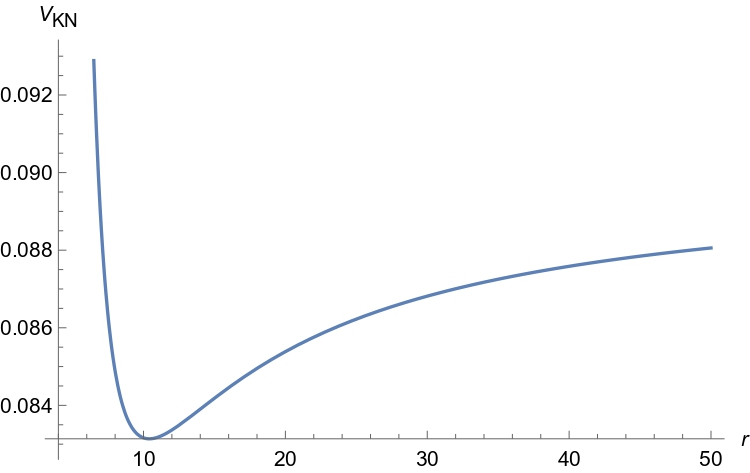}
 \caption[]{The effective potential $V_{KN}(r)$ has a trapping well outside the horizon. So the system is superradiantly unstable. The parameters of the perturbation are taken as $m=1$, $\omega=0.286$, $\mu=0.300$ and $q=0.300$. The parameters of the black hole are $a=2 $, $M=3$, $Q=-1$, $\beta=2.000$.}
 \label{figure:4}
 \end{center}
\end{figure} 
\item\textit{Case-2}. In this case we choose our parameters such that they do not satisfy all the (above mentioned) inequalities.
For example we take $a=2 $, $m=1$, $\omega=0.286$, $\mu=0.300$, $M=3$, $Q=-1$, $q=0.300$, $\beta=2.000$. In this case both bound state condition \ref{exponentially-bound} and superradiance condition \ref{superradiance condition-kerr-newman} hold, but inequality \ref{condition-at-infiniy-KN} does not hold. As evident from the \ref{figure:4}, in this case we have a potential well outside the horizon and the system is superradiantly unstable.
\end{enumerate}
\section{Discussions}\label{section-5}
In the above analysis, we have found superradiantly stable parameter regime of braneworld extremal Kerr and Kerr-Newman black holes under massive scalar perturbation. In our calculations, we have assumed that both the superradiance condition and the bound state condition hold simultaneously. For superradiantly stable parameter space, there should not be any trapping well of the effective potential outside the horizon. Otherwise, the trapped modes will be superradiantly amplified, this leads to the instability of the system. It is the crucial point of our analysis. The radial part of the Klein-Gordon equation is the main object of our concern. After choosing a suitable variable we have reduced the radial equation in the Schr\"odinger form. From this equation, we have determined the effective potential felt by the scalar field. Next, one should choose the parameters in a way such that there remain only one maxima (and no local minima) of the effective potential outside the horizon. This requirement is implemented by demanding that asymptotically the effective potential should behave like decreasing function and there should remain only one positive root of derivative of effective potential outside the horizon. Using this argument and Vieta's theorem, we have obtained some inequalities involving both black hole parameters and perturbation parameters. These inequalities together with the superradiance condition and bound state condition specify the parameter space of stability. For massive scalar perturbation of braneworld Kerr-black holes, we have obtained a bound on tidal charge. On the other hand for charged massive scalar perturbation of slowly rotating braneworld extremal Kerr-Newman black holes, we have obtained a lower bound of the product of mass to charge ratio of black hole and scalar field.  
\section*{Acknowledegment}
I thank Sumanta Chakraborty for giving me the correct suggestion about the metric used in \ref{section-4}. I also thank the unknown reviewer for making some valuable comments about this. My research is supported by the graduate fellowship of IACS. I thank Ramit Dey for a little help in latex.

\bibliographystyle{unsrt}

\bibliography{Extremal-brane.bib}

\end{document}